\documentclass[a4paper]{article}

\usepackage{amsmath}
\usepackage{amssymb}
\usepackage{icrc2013}
\usepackage[none]{hyphenat}
\title{Boosting the performance of the ASTRI SST-2M prototype: reflective and anti-reflective coatings.}

\shorttitle{Optical coatings for ASTRI SST-2M}

\authors{
Giacomo Bonnoli$^{1}$,
Rodolfo Canestrari$^{1}$,
Osvaldo Catalano$^{2}$,
Giovanni Pareschi$^{1}$
 and Luca Stringhetti$^{3}$
for the ASTRI Collaboration, Luca Perri$^{1}$.
}

\afiliations{
$^1$ INAF - Osservatorio Astronomico di Brera, Via E. Bianchi 46, 23807
  Merate, Italy\\
$^{2}$ INAF - Istituto di Astrofisica Spaziale e Fisica Cosmica di Palermo, Via U. La Malfa 153, 90146 Palermo, Italy\\
$^{3}$ INAF - Istituto di Astrofisica Spaziale e Fisica Cosmica di Milano, Via E. Bassini 15, 20133 Milano, Italy\\
}

\email{giacomo.bonnoli@brera.inaf.it}

\abstract{ASTRI is a Flagship Project of the Italian Ministry of Education, University and Research, led by the Italian National Institute of Astrophysics, INAF.  
One of the main aims of the ASTRI Project is the design, construction and
verification on-field of a dual mirror (2M) end-to-end prototype for the Small Size Telescope (SST) envisaged to become part of the
Cherenkov Telescope Array. The ASTRI SST-2M  prototype adopts the 
Schwarzschild-Couder design, and a camera based on SiPM (Silicon
Photo Multiplier); it will be assembled at the
INAF astronomical site of Serra La Nave on mount Etna (Catania, Italy)
within mid 2014, and will start scientific validation phase soon after. The peculiarities of the optical design and of the SiPM
bandpass pushed towards specifically optimized choices in terms of reflective coatings
for both the primary and the secondary mirror. In particular,
multi-layer dielectric coatings, capable of filtering out the large Night Sky Background
contamination at wavelengths $\lambda \gtrsim 700$ nm have been developed and
tested, as a solution for the primary mirrors. Due to the conformation of the
ASTRI SST-2M camera, a reimaging system based on thin pyramidal
light guides could be optionally integrated aiming to increase
the fill factor. An anti-reflective coating optimized for a wide
range of incident angles faraway from normality was specifically developed to
enhance the UV-optical transparency of these elements. The issues, strategy, simulations and experimental results are thoroughly presented.}

\keywords{ mirrors, interferential coatings, ASTRI, CTA, TeV astronomy, IACT.}

\begin{document}

\maketitle

%Begin a section.
\section{Introduction}

 ASTRI ("Astrofisica con Specchi a Tecnologia Replicante Italiana", \cite{bib:pareschiICRC13}) is a flagship project
of the Italian Ministry of Education, University and Research (MIUR) strictly linked to the
development of the ambitious Cherenkov Telescope Array (CTA) \cite{bib:actis11,bib:acharya13}. CTA plans the
construction of many tens of telescopes divided in three kinds of configurations, in
order to cover the energy range from tens of GeV (Large Size Telescope), to
tens of TeV (Medium Size Telescope), and up to 100 TeV (Small Size
Telescope, SST).
Within this framework INAF, the Italian National Institute of Astrophysics,  is currently developing an end-to-end prototype of the CTA
SST in a dual-mirror configuration \nohyphens{(SST-2M)} to be tested under field
conditions \cite{bib:astriExAs}. For the first time, a wide (semiaperture 4.8$^\circ$)  field of view dual-mirror
Schwarzschild-Couder (SC) optical design (proposed already in
1905 but never applied in astronomy yet, see e.g. \cite{bib:vassiliev08}) will
be adopted on a Cherenkov telescope, 
in order to obtain a compact (f-number f/0.5) optical
configuration and small aberrations across the
whole field of view\cite{bib:canestrariICRC13}. Moreover the prototype will be equipped with a light  and compact camera
\cite{bib:catalanoICRC13} based on Silicon Photo Multipliers (SiPM) instead of
the photomultiplier tubes (PMT) that have been extensively adopted so far; this
is a viable (as shown by FACT \cite{bib:anderhub11}) and promising solution, because \nohyphens{SiPM} avoid some of the disadvantages of PMT: namely the  large dimensions, weight and power consumption.
The ASTRI SST-2M prototype will be placed at Serra La Nave, 1735 m a.s.l. on the Etna
Mountain \nohyphens{near} Catania (Italy) at the INAF "M.G. Fracastoro" observing station\cite{bib:maccaroneICRC13};  the
data acquisition is scheduled to start in 2014. Detailed information on all
the aspects of the prototype can be found in these proceedings. In this paper we will discuss the
development of thin-film coatings for the mirrors and the reimaging system of
the ASTRI SST-2M prototype. The peculiarities of the ASTRI SST-2M prototype in terms of optical design and
sensors led to original issues
regarding the reflective coatings to be adopted for the elements of the optical
chain, that will be addressed in the following sections. 
More specifically the design requires three different coatings: two high
reflectance coatings for the primary and secondary mirrors: the planned
solution for these are presented in
 section \ref{HIGHREF}.
Moreover the design of the \nohyphens{ASTRI} SST-2M prototype comprises one anti-reflective coating, to be deposed on the front surface of the
light guides that could be adopted to enhance the fill factor of the SiPM array
covering the focal surface. This development will be addressed in section
\ref{LOWREF}. A fourth element involved in the optical path is the water-tight curved
plexiglass cover protecting the focal surface. This element also will need a
trasmission-enhancing coating but this development will be treated elsewhere.

\section{High reflectivity coatings for the mirrors}
\label{HIGHREF}
 The mirrors for the ASTRI SST-2M prototype will be produced with
  adaptations of the  cold shaping
method developed by the Brera Astronomical Observatory \cite{bib:vernani08,bib:canestrari12}. Details on
the optical design  and on the structure and
production of the mirrors can be found in
\cite{bib:canestrariICRC13}.  The baseline for the mirror coatings of the prototype is represented by
  the well proven $Al$+$SiO_2$ design that is already in operation, for
  instance, on 100
m$^2$ of panels built with the same technology adopted
here\cite{bib:pareschi08} and installed on the \nohyphens{MAGIC-II} reflector since 2009 \cite{bib:doro08}. 
 This coating design grants \nohyphens{all} the needed characteristics both in terms of performance and
durability, and has been robustly proven along years of scientific operation in an hazardous mountain environment.
Given the novelty of the design of the ASTRI \nohyphens{SST-2M} prototype,
we accordingly studied new solutions for the coatings.
 The main specificity of the ASTRI SST-2M design, driving this work, is the choice of
SiPM as sensors instead of the
long-standing standard constituted by PMT.
The photon detection efficiency (PDE) of the SiPM adopted for the prototype
has a non negligible tail above 700 nm and up to 900 nm. This implies a strong
contamination by intense molecular bands of the Night Sky Background light
(NSB) in a
region where the Cherenkov signal is relatively dim. This pushes towards the
introduction of a low-pass filter in the optical chain, achievable by means of interferential coatings deposed onto glass substrates, for
instance, by means of physical vapour deposition (PVD, see
e.g. \cite{bib:macleod}) of dielectric materials such as
  $SiO_2$, $TiO_2$, $Ta_20_5$, $MgF_2$ and many others.  This approach  is
widely investigated  within the CTA Collaboration \cite{bib:bonardiICRC11,bib:foersterICRC13} also because many of
 the eligible materials are expected to adhere to glass better than aluminium,
thus improving durability and resistance to environmental \nohyphens{hazards}.
In this technique  multiple layers of materials characterized by different refractive
indexes, are deposed onto the glass substrate with uniform thicknesses ranging from tens to
hundreds of nanometers.  A proper tuning of
the \nohyphens{materials} and of the layer thicknesses can approach closely an ideal
bandpass filter, with transmittance (or reflectivity) approaching 100\% in the
band of interest and very small \nohyphens{values} elsewhere. These  optical properties
arise from the interference of incident and reflected rays at each
layer interface, therefore the optical behaviour is
severely dependent from the angle of incidence (AoI). 
 From the technical point of view, the preferential site for installing this
 filter would be the focal surface, because it would allow to coat substrates
 of very \nohyphens{small} dimensions (see figure \ref{Piram}); but in
 the SC design the focal surface is interested by rays
 spanning a huge (20$^\circ - 70^\circ$) interval of AoI, spoiling the uniformity of
 the filter transmittance.
As a consequence in the case of the ASTRI SST-2M prototype an effective cut can be
obtained only by tuning the reflectivity curve of the primary dish, where AoI
span the range 0-15$^\circ$, while already the secondary
mirror works under incidences in the wide 20$^\circ - 60^\circ$ interval. 

\subsection{Optimization of the bandpass}

On the pathway to design properly an interferential coating for the ASTRI SST-2M
primary mirror, we tried to assess the optimal position of the red cut of the
passband, in order to tune the desired performance of the coating (assumed as
a good approximation of an ideal step function) and hence
its design. 
We evaluated under very simple assumptions the impact of a low-pass wavelength
cut onto the signal-to-noise ratio (SNR) obtained in a pixel interested by
NSB and a Cherenkov signal. In the background limited case,
and assuming that NSB is Poisson distributed around its average rate $B$,
the excess $S$ due to Cherenkov has:

\begin{equation}
SNR(\lambda)=\frac{S(\lambda)}{\sqrt{B(\lambda)}}
\end{equation}
where $\lambda$ is the long wavelength cut that is meant to be optimized, and
$S(\lambda)$ and $B(\lambda)$ are the integrated counts up to the wavelength $\lambda$.
The maximum condition is therefore

\begin{equation}
\frac{dS(\lambda)}{S(\lambda)}=\frac{dB(\lambda)}{2B(\lambda)}
\end{equation}

and is independent by the intensity of the signal relative to
background under these assumptions, but  \nohyphens{depends} only on the spectral shape of
the Cherenkov signal and of the NSB, and on the PDE curve. Adopting  NSB and Cherenkov spectra taken from literature, and a PDE
interpolated from measurements performed within the activities described in
\cite{bib:catalanoICRC13} we computed that the maximum $SNR$ is reached integrating
signal up to $\sim500$ nm. Extending the band redwards reduces progressively the SNR
but the drop is less than 10\% up to 700 nm, allowing a tolerance useful to
match the constraints cast onto the development of the real coating by the
limited choice of suitable materials and other technical and industrial issues.
Another constraint to keep into account is that given the small ($\sim 7$ m$^2$) effective area
of the pupil, the design must at the same time tend to preserve a large
fraction (e.g. 80\%) of the signal photons detectable by SiPM. 
Moreover, especially for the large signals preferentially associated to multi-TeV showers
that are the core target of SST, the width of the passband increases the signal
photon statistics and hence the precision in the measurement of the integrated
light of the shower and eventually of the energy of the primary.  Eventually,
the performance is meant to meet the CTA requirement for mirror reflectivity,
implying average reflectivity in the $300-550$ nm passband $R^{550}_{300}\ge 83\%$.

\subsection{Coating for the primary dish}

\begin{figure}%[t]
  \centering
 \includegraphics[angle=270,width=0.40\textwidth]{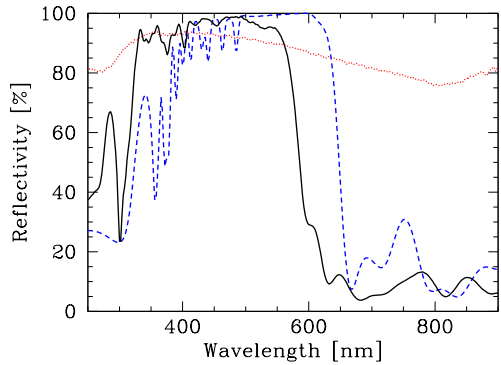}
  \caption{Measured reflectivities for the considered coating designs:
    $Al$+$SiO_2$ (dotted red) $SiO_2$+$TiO_2$ ML (dashed blue) and $SiO_2$+mixture ML (solid black).}
  \label{M1refplot}
 \end{figure}

The primary dish of ASTRI SST-2M is composed of 18 hexagonal panels, each 85 cm wide (face to
face; radius of the circumscribed circle $\sim 98$ cm), arranged in three
concentric rings \cite{bib:canestrariICRC13}. The radii of curvature (RoC) are \nohyphens{different} for
each ring, ranging from $\sim$ 8 m for the inner ones up to  $\sim$ 10 m for the outer
ring, with a sagitta  $S \simeq 11-13$ mm. The relative flatness of
the surface and moderate dimensions of each panel make the primary dish the
preferential locus for application of an interferential filter made of
a multi-layer (ML) coating, from an applicative point of view. This converges with
the more fundamental constraint that in the adopted design the primary
dish is the only optical element interested by rays incoming with AoI within a
narrow range around normality (up to 15$^\circ$).
A first test was performed in the  ZAOT S.r.l. laboratories\footnote{{\tt
      http://www.zaot.com}} with a \nohyphens{ML} adopting $SiO_2$ as low index material and $TiO_2$ as high index material.
Sample glass windows where coated along the diagonal of the hexagon in order
to check uniformity across the width of the coating chamber. The reflectivity
curve was then averaged weighting for the surface area at each distance from the center, in order to
obtain the average response of the panel at each wavelength under the
assumption of azimuthal simmetry.
The sharpness of the cut was conserved in the averaged reflectance, proving
that a good spatial homogeneity of the deposition process was achieved, with
the exception of small central areas scarcely affecting the overall behaviour
of the panel. The sample windows were then subject to tape tests, a thermal cycling between
$-25 ^\circ$C and $+60 ^\circ$C and salt mist test, showing no degradation either in adhesion and appearance or in reflectivity.
Two test panels produced with the composite technology described in
\cite{bib:canestrariICRC13} have already been coated, one with an $Al$+$SiO_2$ recipe and one with the $SiO_2$+$TiO_2$ ML coating. These
panels have been exposed outdoors in the park of the Merate site of the Brera
Observatory, as a comparative test for resistance to environmental hazards and
mirror misting; the latter test is important as recent studies hint to dielectric
  coated mirrors being prone to condensation \cite{bib:chadwickICRC13}. 
 The ML performance in near UV is limited because $TiO_2$ causes
significant absorption below 380 nm. A further improvement was then seeked 
substituting $TiO_2$  with  a dioxide mixture  with refractive index $n\simeq2.1$. Again
samples disposed along the diagonal of a panel were coated, the reflectivity
measured and the spatial evolution of reflectance interpolated and
area-averaged. This solution showed a better behaviour at wavelengths below
400 nm and is, together with the baseline coating represented by an $Al$+$SiO_2$
design, compliant with the current CTA requirement of an average
reflectivity of at least 83\% in the $300-550$ nm range. The measured
reflectivities for the three designs are plotted in figure \ref{M1refplot}.

\subsection{Coating for the secondary mirror}

\begin{figure}%[t]
  \centering
 \includegraphics[angle=270,width=0.40\textwidth]{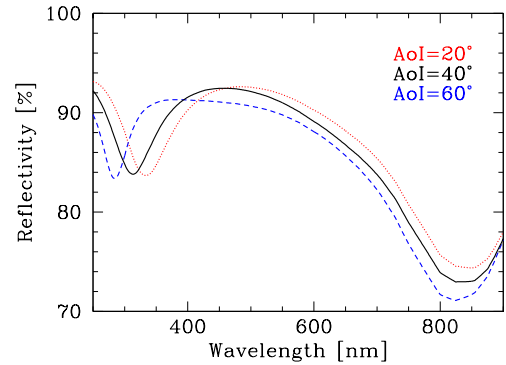}
  \caption{Simulated reflectivity for the simple $Al$+$SiO_2$ design optimized for
    the ASTRI SST-2M secondary mirror. Curves for incidence angles of
    20$^\circ$ (dotted red), 40$^\circ$ (solid black), and
  60$^\circ$ (dashed blue) are plotted.}
  \label{M2ref}
 \end{figure}

The secondary mirror of ASTRI SST-2M consists of a \nohyphens{single panel}, whether the full
glass or composite (see \cite{bib:canestrariICRC13}) solution will be preferred. The large
(1.80 m) diameter, considerable weight ($\sim 120$ kg) and pronounced
curvature ($RoC=2.2$ m, for a sagitta of $\approx 22$ cm) discourage the deposition
of complex coatings. Moreover, in the optical design the distribution of
angles of incidence is wide and far from normality (20$^\circ$ to 60$^\circ$,
peaking at 40$^\circ$). All these constraints together drove towards the
simplest design, an $Al$-$SiO_2$ dual layer. This is a well known solution in
the field, granting high reflectivity at all the wavelengths of interest and a
well-established performance in terms of resistance and durability. This choice has
the additional advantage that, if the composite mirror is the preferred
solution for the secondary, the aluminium-quartz coating can be deposed onto
the assembled panel, while other materials such as the dioxides (or fluorides)
used in ML require high temperatures potentially harmful for the glues.
A dedicated simulation pinpointed a thickness of 150 nm for the $SiO_2$ layer
as the optimal
solution granting the required reflectivity in the blue part of the spectrum across the whole AoI range, as
displayed in figure \ref{M2ref}.
  Deposition tests and sample measurements will start in the second half of 2013
in a coating chamber that our industrial partner ZAOT S.r.l. is developing
specifically for this task and is currently under commissioning. 

\section{Anti-reflective coating for the reimaging system}
\label{LOWREF}
\begin{figure}[h]
  \centering
 \includegraphics[width=0.32\textwidth]{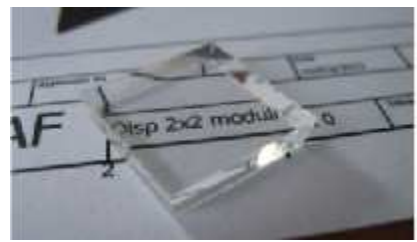}
 \caption{One of the light guides  designed for the focal surface of the ASTRI SST-2M
  prototype.}
  \label{Piram}
 \end{figure}

\begin{figure}%[h]
  \centering
 \includegraphics[width=0.43\textwidth]{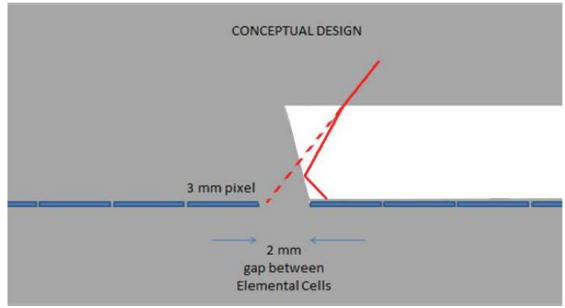}
 \caption{Conceptual design for the reimaging system studied for the ASTRI SST-2M
    prototype. Photons directed onto the dead areas of the focal surface are
    recovered due to the inner total reflection onto the slanted faces of the light guides.}
  \label{Pyr}
 \end{figure}

The Hamamatsu S11828-3344M SiPM units currently available for the ASTRI SST-2M focal surface
are not designed for a tight array arrangement.  Even if future sensors
  will overcome this problem, being based on a different configuration, however
in the case of  the ASTRI SST-2M prototype this introduces significant
dead areas that  could be recovered by a dedicated reimaging
system. The system consists of 14 mm $\times$ 14 mm wide, 2.5 mm thick truncated
pyramids made of LAK9, a high
refractive index ($n=1.69$) glass (see figure \ref{Piram}). Each light guide
 is meant to be glued
onto a sensor unit and recover photons due to the total internal reflection onto its
slanted faces (see figure \ref{Pyr}).

\begin{figure}[t]
  \centering
 \includegraphics[angle=270,width=0.40\textwidth]{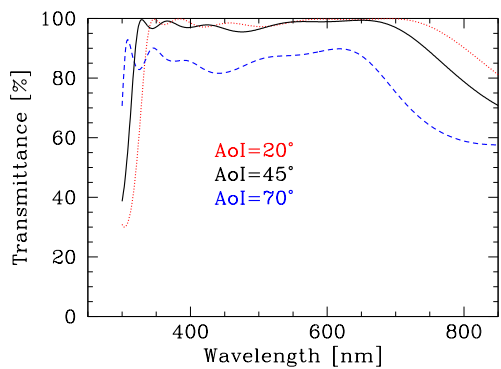}
  \caption{Simulated  transmission curves at the air-glass interface for LAK9
    glass coated with the $SiO_2$+$ZrO_2$ multilayer designed for the ASTRI
    SST-2M reimaging system. Curves for AoI of 20$^\circ$ (dotted red), 45$^\circ$
    (solid black) and 70$^\circ$ (dashed blue) are plotted.}
  \label{Pyr2}
 \end{figure}

\begin{figure}%[t]
  \centering
 \includegraphics[angle=270,width=0.40\textwidth]{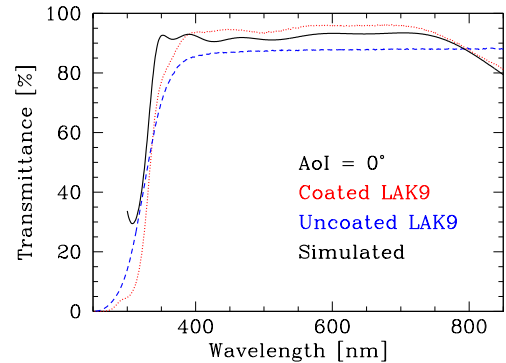}
 \caption{Measured transmission curve for a coated sample of LAK9 glass under
    normal incidence (dotted red). The curve is compared with the uncoated glass
    sample (dashed blue) and the simulated behaviour of the adopted design (solid black). The
    careful reader will note that the simulated performance does not reach 100\%
    transmission as in the curves of figure \ref{Pyr2}. This is partly due
    to the fact that the design is \emph{not} optimized for normal incidences
    and partly because of the exit (glass to air) interface that is
    unavoidable in our measurement setup and was added specifically in the simulation reported here for a correct comparison.}
  \label{Pyr3}
 \end{figure}

 An anti-reflective coating is then needed
to reduce the light loss ($\sim 13\%$) due to the reflection on the bases of the light
guides. This design is challenging as AoI on the focal surface range from
20$^\circ$ to 70$^\circ$, and the bandpass of the coating must match the one
of the filter implemented onto the primary across the whole range. This was obtained (see
figure \ref{Pyr2}) in the simulations with a design including $SiO_2$ and $ZrO_2$ alternate layers.
 A first set of $\sim 80$ light guides was coated by ZAOT S.r.l and the
transmission measured at normal incidence (choice obliged by limitation of our
measurement setup) is shown in figure \ref{Pyr3}, compared
with the uncoated glass and with the simulation.

\section{Conclusions}

We conclude comparing (see table \ref{results}) the different designs produced
so far for the coating of the primary dish, evaluating the effectiveness of the whole system (including
the SiPM PDE)  in terms of acceptance for the Cherenkov signal $A$
and rejection power for NSB $B$. We also define a quality factor $Q\equiv A/(1-B)$.  
We use the measured reflectivity curves for the primary mirror, and the simulated curves for the reflectivity of the
secondary and the transmittance of the reimaging system. The calculation is
performed considering normal incidence on the primary and intermediate AoI on
the secondary (40$^\circ$) and on the focal surface (50$^\circ$).
The average reflectivity in the band $300-550$  $R_{300}^{550}$ is also
reported in the last column. The proposed dielectric
solutions significantly contribute to optimize the performance of the ASTRI
SST-2M prototype when compared to the baseline $Al$+$SiO_2$ design.

\begin{table}[h]
\begin{center}
\begin{tabular}{|l|ccc|c|}
\hline Coating  design& $A$ & $B$ & $Q$ & $R_{300}^{550}$\\
&[\%]&[\%]&&[\%]\\
 \hline
$Al$+$SiO_2$   & 24.0&94.1 & 4.1& 91.9\\ %\hline
$SiO_2$+$TiO_2$ ML & 23.3  & 96.0  & 5.9 &79.7\\ %\hline
$SiO_2$+mixture ML & 21.9  & 96.7 &6.5&90.6\\ %\hline
\hline
\end{tabular}
\caption{Comparison of the different coating designs considered for the
  primary mirror (see text).}
\label{results}
\end{center}
\end{table}

%%%%%%%%%%%%%%%%%%%%%%%%%%%%%%%%%
\vspace*{0.5cm}
\footnotesize{{\bf Acknowledgment:}{ This work was partially supported by the
    ASTRI “Flagship Project” financed by the Italian Ministry of Education,
    University, and Research (MIUR) and led by the Italian National Institute
    of Astrophysics (INAF). We also acknowledge partial support by the MIUR
    ``Bando PRIN 2009''. The authors are grateful to ZAOT S.r.l. and in
    particular Mr. Adelfio Zanoni for the significant expertise provided and
    the fruitful collaborative efforts devoted to this intriguing task.}}

\end{document}